\documentclass[aps,twocolumn,superscriptaddress]{revtex4}
\usepackage{graphicx}  
\usepackage{dcolumn}   
\usepackage{bm}        
\usepackage{amssymb}   
\usepackage{multirow}
\usepackage{lineno}

\begin{document}


\newcommand{\dzero}     {D0}
\newcommand{\ttbar}     {\mbox{$t\bar{t}$}}
\newcommand{\bbbar}     {\mbox{$b\bar{b}$}}
\newcommand{\ccbar}     {\mbox{$c\bar{c}$}}
\newcommand{\herwig}    {{\sc{herwig}}}
\newcommand{\pythia}    {{\sc{pythia}}}
\newcommand{\vecbos}    {{\sc{vecbos}}}
\newcommand{\alpgen}    {{\sc{alpgen}}}
\newcommand{\qq}        {\sc{qq}}
\newcommand{\evtgen}    {{\sc{evtgen}}}
\newcommand{\tauola}    {{\sc{tauola}}}
\newcommand{\geant}     {{\sc{geant}}}
\newcommand{\metcal}    {\mbox{$\not\!\!E_{Tcal}$}}
\newcommand{\met}       {\mbox{$\not\!\!E_T$}}
\newcommand{\pt}        {$p_T$}
\newcommand{\sigmatt}   {\ensuremath{\sigma_{t\bar{t}}}}
\newcommand{\mtop}      {$m_{\rm top}$}
\newcommand{\rsigma}    {\ensuremath{R_{\sigma}}}

\newcommand{\feynhiggs} {{\sc{FeynHiggs}}}
\newcommand{\cpsuperh} {{\sc{CPsuperH}}}

\newcommand{\brh}{\ensuremath{{ B}(t\rightarrow H^+ b)}}

\newcommand{\lumi}      {1~$\rm fb^{-1}$}
\newcommand{\result}    {X.X}
\newcommand{\erstat}    {^{+X.X}_{-X.X}}
\newcommand{\ersyspos}  {+X.X}
\newcommand{\ersysneg}  {-X.X}
\newcommand{\ersys}     {^{\ersyspos}_{\ersysneg}}
\newcommand{\erlumi}    {X.X}
\newcommand{\ljets}     {\ensuremath{\ell}+\rm{jets}}
\newcommand{\eplus}     {\ensuremath{e}+jets}
\newcommand{\muplus}    {\ensuremath{\mu}+jets}
\newcommand{\ltau}      {\ensuremath{\tau \ell}}
\newcommand{\etau}      {\ensuremath{\tau e}}
\newcommand{\mutau}     {\ensuremath{\tau \mu}}

\newcommand{\rs}        {XX.X}
\newcommand{\rserr}     {^{+X.X}_{-X.X}}
\newcommand{\rsigmaell}    {\ensuremath{R_{\ell\ell/\ell j}}}
\newcommand{\rsigmatau}    {\ensuremath{R_{\ltau/\ell \ell\text{-}\ell j }}}


\title{Measurement of $d\sigma/dy$ of Drell-Yan $e^{+}e^{-}$ pairs in the $Z$ Mass Region from 
$p\bar{p}$ Collisions at $\sqrt{s}$=1.96 TeV}

\affiliation{Institute of Physics, Academia Sinica, Taipei, Taiwan 11529, Republic of China} 
\affiliation{Argonne National Laboratory, Argonne, Illinois 60439, USA} 
\affiliation{University of Athens, 157 71 Athens, Greece} 
\affiliation{Institut de Fisica d'Altes Energies, Universitat Autonoma de Barcelona, E-08193, Bellaterra (Barcelona), Spain} 
\affiliation{Baylor University, Waco, Texas 76798, USA} 
\affiliation{Istituto Nazionale di Fisica Nucleare Bologna, $^{dd}$University of Bologna, I-40127 Bologna, Italy} 
\affiliation{Brandeis University, Waltham, Massachusetts 02254, USA} 
\affiliation{University of California, Davis, Davis, California 95616, USA} 
\affiliation{University of California, Los Angeles, Los Angeles, California 90024, USA} 
\affiliation{University of California, San Diego, La Jolla, California 92093, USA} 
\affiliation{University of California, Santa Barbara, Santa Barbara, California 93106, USA} 
\affiliation{Instituto de Fisica de Cantabria, CSIC-University of Cantabria, 39005 Santander, Spain} 
\affiliation{Carnegie Mellon University, Pittsburgh, Pennsylvania 15213, USA} 
\affiliation{Enrico Fermi Institute, University of Chicago, Chicago, Illinois 60637, USA}
\affiliation{Comenius University, 842 48 Bratislava, Slovakia; Institute of Experimental Physics, 040 01 Kosice, Slovakia} 
\affiliation{Joint Institute for Nuclear Research, RU-141980 Dubna, Russia} 
\affiliation{Duke University, Durham, North Carolina 27708, USA} 
\affiliation{Fermi National Accelerator Laboratory, Batavia, Illinois 60510, USA} 
\affiliation{University of Florida, Gainesville, Florida 32611, USA} 
\affiliation{Laboratori Nazionali di Frascati, Istituto Nazionale di Fisica Nucleare, I-00044 Frascati, Italy} 
\affiliation{University of Geneva, CH-1211 Geneva 4, Switzerland} 
\affiliation{Glasgow University, Glasgow G12 8QQ, United Kingdom} 
\affiliation{Harvard University, Cambridge, Massachusetts 02138, USA} 
\affiliation{Division of High Energy Physics, Department of Physics, University of Helsinki and Helsinki Institute of Physics, FIN-00014, Helsinki, Finland} 
\affiliation{University of Illinois, Urbana, Illinois 61801, USA} 
\affiliation{The Johns Hopkins University, Baltimore, Maryland 21218, USA} 
\affiliation{Institut f\"{u}r Experimentelle Kernphysik, Karlsruhe Institute of Technology, D-76131 Karlsruhe, Germany} 
\affiliation{Center for High Energy Physics: Kyungpook National University, Daegu 702-701, Korea; Seoul National University, Seoul 151-742, Korea; Sungkyunkwan University, Suwon 440-746, Korea; Korea Institute of Science and Technology Information, Daejeon 305-806, Korea; Chonnam National University, Gwangju 500-757, Korea; Chonbuk National University, Jeonju 561-756, Korea} 
\affiliation{Ernest Orlando Lawrence Berkeley National Laboratory, Berkeley, California 94720, USA} 
\affiliation{University of Liverpool, Liverpool L69 7ZE, United Kingdom} 
\affiliation{University College London, London WC1E 6BT, United Kingdom} 
\affiliation{Centro de Investigaciones Energeticas Medioambientales y Tecnologicas, E-28040 Madrid, Spain} 
\affiliation{Massachusetts Institute of Technology, Cambridge, Massachusetts 02139, USA} 
\affiliation{Institute of Particle Physics: McGill University, Montr\'{e}al, Qu\'{e}bec, Canada H3A~2T8; Simon Fraser University, Burnaby, British Columbia, Canada V5A~1S6; University of Toronto, Toronto, Ontario, Canada M5S~1A7; and TRIUMF, Vancouver, British Columbia, Canada V6T~2A3} 
\affiliation{University of Michigan, Ann Arbor, Michigan 48109, USA} 
\affiliation{Michigan State University, East Lansing, Michigan 48824, USA}
\affiliation{Institution for Theoretical and Experimental Physics, ITEP, Moscow 117259, Russia} 
\affiliation{University of New Mexico, Albuquerque, New Mexico 87131, USA} 
\affiliation{Northwestern University, Evanston, Illinois 60208, USA} 
\affiliation{The Ohio State University, Columbus, Ohio 43210, USA} 
\affiliation{Okayama University, Okayama 700-8530, Japan} 
\affiliation{Osaka City University, Osaka 588, Japan} 
\affiliation{University of Oxford, Oxford OX1 3RH, United Kingdom} 
\affiliation{Istituto Nazionale di Fisica Nucleare, Sezione di Padova-Trento, $^{ee}$University of Padova, I-35131 Padova, Italy} 
\affiliation{LPNHE, Universite Pierre et Marie Curie/IN2P3-CNRS, UMR7585, Paris, F-75252 France} 
\affiliation{University of Pennsylvania, Philadelphia, Pennsylvania 19104, USA}
\affiliation{Istituto Nazionale di Fisica Nucleare Pisa, $^{ff}$University of Pisa, $^{gg}$University of Siena and $^{hh}$Scuola Normale Superiore, I-56127 Pisa, Italy} 
\affiliation{University of Pittsburgh, Pittsburgh, Pennsylvania 15260, USA} 
\affiliation{Purdue University, West Lafayette, Indiana 47907, USA} 
\affiliation{University of Rochester, Rochester, New York 14627, USA} 
\affiliation{The Rockefeller University, New York, New York 10021, USA} 
\affiliation{Istituto Nazionale di Fisica Nucleare, Sezione di Roma 1, $^{ii}$Sapienza Universit\`{a} di Roma, I-00185 Roma, Italy} 

\affiliation{Rutgers University, Piscataway, New Jersey 08855, USA} 
\affiliation{Texas A\&M University, College Station, Texas 77843, USA} 
\affiliation{Istituto Nazionale di Fisica Nucleare Trieste/Udine, I-34100 Trieste, $^{jj}$University of Trieste/Udine, I-33100 Udine, Italy} 
\affiliation{University of Tsukuba, Tsukuba, Ibaraki 305, Japan} 
\affiliation{Tufts University, Medford, Massachusetts 02155, USA} 
\affiliation{Waseda University, Tokyo 169, Japan} 
\affiliation{Wayne State University, Detroit, Michigan 48201, USA} 
\affiliation{University of Wisconsin, Madison, Wisconsin 53706, USA} 
\affiliation{Yale University, New Haven, Connecticut 06520, USA} 
\author{T.~Aaltonen}
\affiliation{Division of High Energy Physics, Department of Physics, University of Helsinki and Helsinki Institute of Physics, FIN-00014, Helsinki, Finland}
\author{J.~Adelman}
\affiliation{Enrico Fermi Institute, University of Chicago, Chicago, Illinois 60637, USA}
\author{B.~\'{A}lvarez~Gonz\'{a}lez$^w$}
\affiliation{Instituto de Fisica de Cantabria, CSIC-University of Cantabria, 39005 Santander, Spain}
\author{S.~Amerio$^{ee}$}
\affiliation{Istituto Nazionale di Fisica Nucleare, Sezione di Padova-Trento, $^{ee}$University of Padova, I-35131 Padova, Italy} 

\author{D.~Amidei}
\affiliation{University of Michigan, Ann Arbor, Michigan 48109, USA}
\author{A.~Anastassov}
\affiliation{Northwestern University, Evanston, Illinois 60208, USA}
\author{A.~Annovi}
\affiliation{Laboratori Nazionali di Frascati, Istituto Nazionale di Fisica Nucleare, I-00044 Frascati, Italy}
\author{J.~Antos}
\affiliation{Comenius University, 842 48 Bratislava, Slovakia; Institute of Experimental Physics, 040 01 Kosice, Slovakia}
\author{G.~Apollinari}
\affiliation{Fermi National Accelerator Laboratory, Batavia, Illinois 60510, USA}
\author{J.~Appel}
\affiliation{Fermi National Accelerator Laboratory, Batavia, Illinois 60510, USA}
% ...

\collaboration{CDF Collaboration\footnote{With visitors from $^a$University of Massachusetts Amherst, Amherst, Massachusetts 01003,
$^b$Universiteit Antwerpen, B-2610 Antwerp, Belgium, 
$^c$University of Bristol, Bristol BS8 1TL, United Kingdom,
$^d$Chinese Academy of Sciences, Beijing 100864, China, 
$^e$Istituto Nazionale di Fisica Nucleare, Sezione di Cagliari, 09042 Monserrato (Cagliari), Italy,
$^f$University of California Irvine, Irvine, CA  92697, 
$^g$University of California Santa Cruz, Santa Cruz, CA  95064, 
$^h$Cornell University, Ithaca, NY  14853, 
$^i$University of Cyprus, Nicosia CY-1678, Cyprus, 
$^j$University College Dublin, Dublin 4, Ireland,
$^k$University of Edinburgh, Edinburgh EH9 3JZ, United Kingdom, 
$^l$University of Fukui, Fukui City, Fukui Prefecture, Japan 910-0017,
$^m$Kinki University, Higashi-Osaka City, Japan 577-8502,
$^n$Universidad Iberoamericana, Mexico D.F., Mexico,
$^o$University of Iowa, Iowa City, IA  52242,
$^p$Kansas State University, Manhattan, KS 66506,
$^q$Queen Mary, University of London, London, E1 4NS, England,
$^r$University of Manchester, Manchester M13 9PL, England,
$^s$Muons, Inc., Batavia, IL 60510, 
$^t$Nagasaki Institute of Applied Science, Nagasaki, Japan, 
$^u$University of Notre Dame, Notre Dame, IN 46556,
$^v$Obninsk State University, Obninsk, Russia,
$^w$University de Oviedo, E-33007 Oviedo, Spain, 
$^x$Texas Tech University, Lubbock, TX  79609, 
$^y$IFIC(CSIC-Universitat de Valencia), 56071 Valencia, Spain,
$^z$Universidad Tecnica Federico Santa Maria, 110v Valparaiso, Chile,
$^{aa}$University of Virginia, Charlottesville, VA  22906,
$^{bb}$Bergische Universit\"at Wuppertal, 42097 Wuppertal, Germany,
$^{cc}$Yarmouk University, Irbid 211-63, Jordan,
$^{kk}$On leave from J.~Stefan Institute, Ljubljana, Slovenia, 
}}
\noaffiliation



\begin{abstract}
We report on a CDF measurement of the total cross section and rapidity distribution,
$d\sigma/dy$, for $q\bar{q}\to \gamma^{*}/Z\to e^{+}e^{-}$ events
in the $Z$ boson mass region ($66<M_{ee}<116$~GeV/c$^2$)
produced in $p\bar{p}$ collisions at $\sqrt{s}=1.96$~TeV with $2.1$~fb$^{-1}$ of integrated luminosity.
The measured cross section of $257\pm16$~pb and $d\sigma/dy$ distribution 
are compared with Next-to-Leading-Order(NLO)
and Next-to-Next-to-Leading-Order(NNLO)
QCD theory predictions with CTEQ
and MRST/MSTW parton distribution functions (PDFs).
There is good agreement between  the experimental  total cross section
and $d\sigma/dy$  measurements with theoretical calculations with the most recent  NNLO PDFs.
\end{abstract}

\maketitle 

\section{Introduction} 
Accurate predictions using perturbative quantum chromodynamics (QCD) are critical
for understanding experimental results at hadron colliders.
Such predictions depend on the accuracy 
of input parton distribution functions (PDFs), which at present 
cannot be calculated and are obtained from analysis of 
data from a broad range of  processes.
Precise knowledge of PDFs will be  particularly 
important for analysis of data at the Large Hadron Collider (LHC)
where new phenomena may be revealed via small  
deviations from Standard Model (SM) predictions.
The Drell-Yan process \cite{drellyan}, in which quark-antiquark annihilations form
intermediate $\gamma^{*}$ or $Z$ ($\gamma^{*}/Z$) vector bosons decaying to
lepton pairs, is particularly useful in providing information on PDFs
at $Q^{2}=M_{\ell\ell}^{2}$, where $M_{\ell\ell}$ is the invariant mass of the dilepton pair. 
In the leading order (LO) approximation, the momentum fractions $x_1$, $x_2$ carried by the initial state quarks
in the proton and antiproton, respectively, are related to the rapidity $y$ \cite{cdf_coord}
of the $\gamma^{*}/Z$ boson via the equation $x_{1,2} = (M_{\ell\ell}/\sqrt{s})e^{\pm y}$, where $\sqrt{s}$ 
is the center of mass energy.
Dilepton pairs produced at large $y$ originate from collisions in which
one parton carries a large and the other a small momentum fraction $x$.
A measurement of $d\sigma/dy$ at large $y$
tests PDFs at high $x$, a region not well constrained by current results. 
Therefore, precise measurements of $W$ and $Z$ boson rapidity distributions at the Tevatron determine
the size of higher order QCD terms and can be used to further refine current PDF models.
Furthermore since the $Z$ production cross section is predicted with an accuracy of $\approx$ 2$\%$ {\cite{mrst2008pdfs}, 
 precise measurements of the rate of $Z$ production at the Tevatron and the LHC can be used to determine the integrated luminosity~{\cite{znorm} more precisely
 than the traditional method of using the total inelastic cross section.
 This has particular applicability to sub-processes initiated by a quark and an anti-quark and can reduce the uncertainty in the
determination of LHC and Tevatron cross sections.

The most recent Tevatron  measurement of $d\sigma/dy$ for $e^{+}e^{-}$
pairs  in the $Z$ boson mass region was performed by the D0 \cite{d0rsl} experiment, using a data-set 
corresponding to $0.4$ fb$^{-1}$ of integrated luminosity.
Here, we report on a new measurement of $d\sigma/dy$ at the Tevatron
with an integrated luminosity of   $2.1$ fb$^{-1}$. 
The measured rapidity range extends to $|y|\sim 2.9$, close to  the kinematic limit  of $|y|=3.0$ for $Z$ boson
production at $\sqrt{s}$~=~1.96 TeV.
The $d\sigma/dy$ distribution is compared to the predictions of perturbative QCD calculations in
Next-to-Leading-Order (NLO)
and Next-to-Next-to-Leading-Order (NNLO) with different PDF models.  

\section{Event selection and analysis method} \label{sec:eventsel}
The data sample  corresponds to an integrated luminosity of 
 $2.1$ fb$^{-1}$ collected by the CDF II Detector
 at Fermilab \cite{cdfdet} during 2004-2007.
CDF II uses a 1.4 T solenoidal magnetic spectrometer surrounded by projective-tower-geometry calorimeters
and outer muon detectors.
Charged particle directions and momenta are measured by an open-cell drift
chamber (COT), a silicon vertex detector (SVX), and an intermediate silicon layer (ISL).
The coverage of COT tracking in pseudorapidity is $|\eta|<1.2$ \cite{cdf_coord}. 
Reconstructed tracks are used to determine 
the $p\bar{p}$ collision point along the beam line ($z_{vertex}$), 
which is required to be within $z=\pm60$ cm of the detector.
The energies and directions \cite{cdf_coord} of electrons, photons,
and jets are measured by two separate calorimeters:
central ($|\eta|<1.1$) and plug ($1.1<|\eta|<3.6$).
Each calorimeter  has an electromagnetic (EM)  compartment with a
shower maximum detector followed by a hadronic (HAD) compartment.
Three topologies of $e^{+}e^{-}$ pairs are considered: two central electrons (CC),
 one central and one plug electron (CP), and two plug electrons (PP).
 The inclusion of PP events allows the measurement of $Z$ bosons in the forward
 rapidity region which corresponds to  high and low parton momentum fractions.

Data are collected using a three-level trigger system \cite{cdfdet} 
and  trigger paths with either one central electron
 or two electrons (central or plug) with  transverse energy $E_{T}>18$ GeV. 
Electron identification requirements \cite{jyhan_thesis} are imposed to select  signal events and to suppress background.
 Both electron candidates are 
required to be isolated from any other calorimetric activity.
The fraction of energy in the HAD calorimeter towers behind the
 EM shower is required to be small \cite{jyhan_thesis}, as expected for an EM shower.
Electron candidates with  $E_{T}>25$ GeV for  
  CC and PP events, and $E_{T}>20$ GeV  for CP events, are selected 
 in the central ($|\eta|<1.1$), and plug  ($1.2<|\eta|<2.8$) 
fiducial regions of the calorimeters.
Central electron candidates must have a COT track that extrapolates to a shower cluster in the EM calorimeter and  a track momentum consistent with the calorimeter measurement.
Central and plug electron candidates are required to have EM-like transverse shower profiles using
the shower maximum detectors.
In order to reduce background we require that 
at least one of the plug electrons in PP events has a track reconstructed in the SVX that
 points to the EM cluster in the calorimeter.
The efficiency of having at least one electron matched to an SVX track is about $85\%$.
 The selected number of CC, CP, and PP events
 with $66<M_{ee}<116$ GeV/c$^{2}$ is 50752, 86203, and 31415, respectively.
\begin{figure}[ht]
\begin{center}
\includegraphics[width=3.0in]{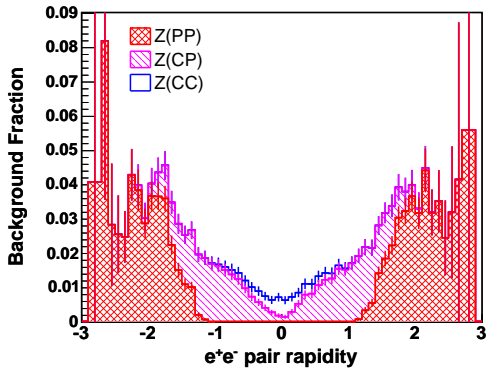}
\end{center}
\caption{The distribution of  background  events as a function of rapidity
for CC, CP and PP $e^{+}e^{-}$ candidates (shown as a fraction of all candidate
events selected in this analysis).
 The error bars include the statistical and systematic uncertainty.}
\label{fig:bkg} 
\end{figure}
 \section{Backgrounds} 
 The main backgrounds are QCD dijet  and photon plus jet events
 in the plug  region (because of the limited  tracking at large $|\eta|$).
 The jet background  is measured separately in each $e^{+}e^{-}$ pair topology 
 by statistically separating electrons from jets on the basis of the transverse energy profile 
 distributions in the calorimeter and  the invariant mass distribution of $e^{+}e^{-}$ pairs \cite{jyhan_thesis}. 
The distribution of  background  events as a function of rapidity
for CC, CP and PP $e^{+}e^{-}$ candidates (shown as a fraction of all candidate
events selected in this analysis) is shown in Fig. \ref{fig:bkg}. 
The fractional contribution of the total background to the number of selected events is 
$0.24\pm0.03\%$ (\textrm{stat} $\oplus$ \textrm{syst}) for CC,
 $1.55\pm 0.44\%$ for CP, and $3.40\pm 0.75\%$ for PP events.
 The  background from electroweak ($WW$, $WZ$, $W$+jets, and $Z\to\tau^{+}\tau^{-}$)
 and $t\bar{t}$ processes is estimated from simulation to be $0.41\pm 0.02\%$.

\section{Acceptance and efficiencies} 
The acceptance is defined as the ratio of the number of Monte Carlo (MC) simulation events that pass selection criteria in each 
 $y$  bin of  the reconstructed final state   $e^{+}e^{-}$ pair (including resolution smearing)  
 to the number of MC generated events in each true $y$ bin of the generated $\gamma^{*}/Z$ boson. 
  The resolution in the measurement of the $e^{+}e^{-}$ invariant mass is 2.2 $GeV/c^2$, 
   and  the resolution in the measurement of $y$ is 0.015.
    The acceptance  is modeled using the {\sc pythia} \cite{pythia} generator
 combined with a {\sc geant} \cite{geant} simulation of the CDF detector.
 
 The  {\sc pythia} generator includes 
  a LO QCD interaction ($q+\bar{q} \to \gamma^{*}/Z$), initial state QCD radiation, parton shower fragmentation, the 
 $\gamma^{*}/Z \to e^{+}e^{-}$ decay, and  photon radiation from the final state. 
 The version of  {\sc pythia}  used at CDF has additional ad-hoc tuning \cite{pythia} in order to accurately represent the
 $Z$ boson transverse momentum distribution measured in data. 
  To reconstruct the simulated events in the same way as data, the calorimetry energy scale, 
  resolutions, and selection efficiencies used in the detector simulation are tuned using data.
 
Because the acceptance depends on modeling of the $Z$  boson rapidity, transverse momentum and angular distributions of the electron pairs, it is important to correct for possible model dependences arising from the choice of the event generator or a particular PDF set. 
  The uncorrected acceptance is calculated using the  CTEQ5L~\cite {cteq5L} LO PDFs, and we compare relevant kinematic distributions in the MC simulation to those observed in the data to correct the acceptance for possible observed discrepancies.
 
While the generated rapidity spectrum is in good agreement with the data for $y < 2$, the data and simulation do not agree at larger values of $y$. To correct for this discrepancy, we modify the MC generated event spectrum ($dN/dy$) so that the
final accepted MC spectrum matches the spectrum in data, as shown in Fig. \ref{fig:yratio}.
A comparison of the reconstructed transverse
momentum spectra of the $e^{+}e^{-}$ pairs in the data and the MC simulation
reveals  good agreement as shown in Fig.~\ref{fig:pt}. 
 Modifying the $P_T$ spectrum in simulation to exactly follow the data leads to a negligible change in the calculated acceptance. A comparison of the average $P_T$ of events in bins of $y$  shows that the data is well modeled by the simulation. 
  Similarly, a study of the angular distribution in $\theta$ (where $\theta$ is the polar angle of the final
state electron in the  Collins-Soper frame~\cite {collins}) in data shows good agreement
with the simulation as illustrated in Fig. ~\ref{fig:angular}.

\begin{figure}[ht]
\begin{center}
 \includegraphics[width=3.0in]{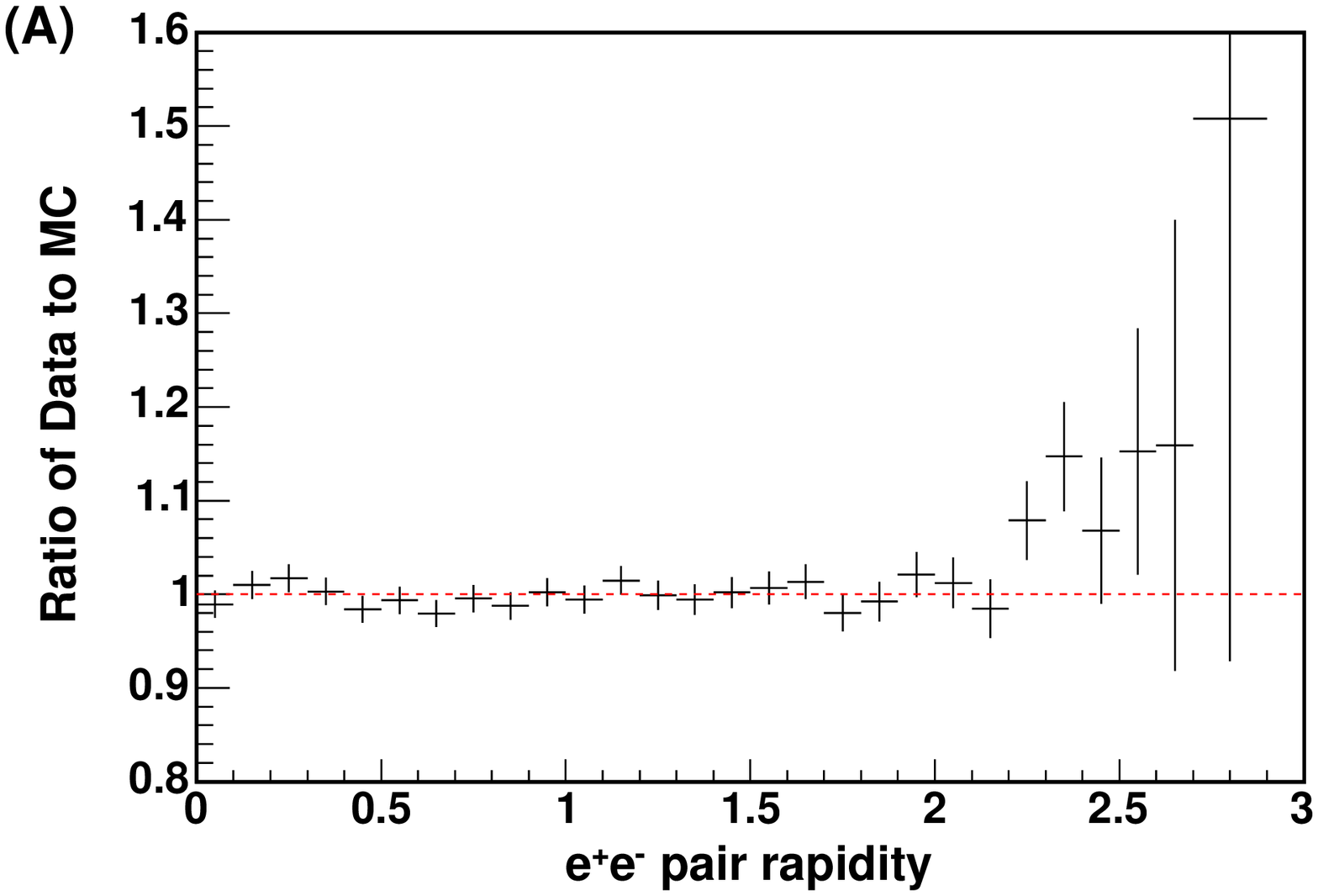}
 \includegraphics[width=3.0in]{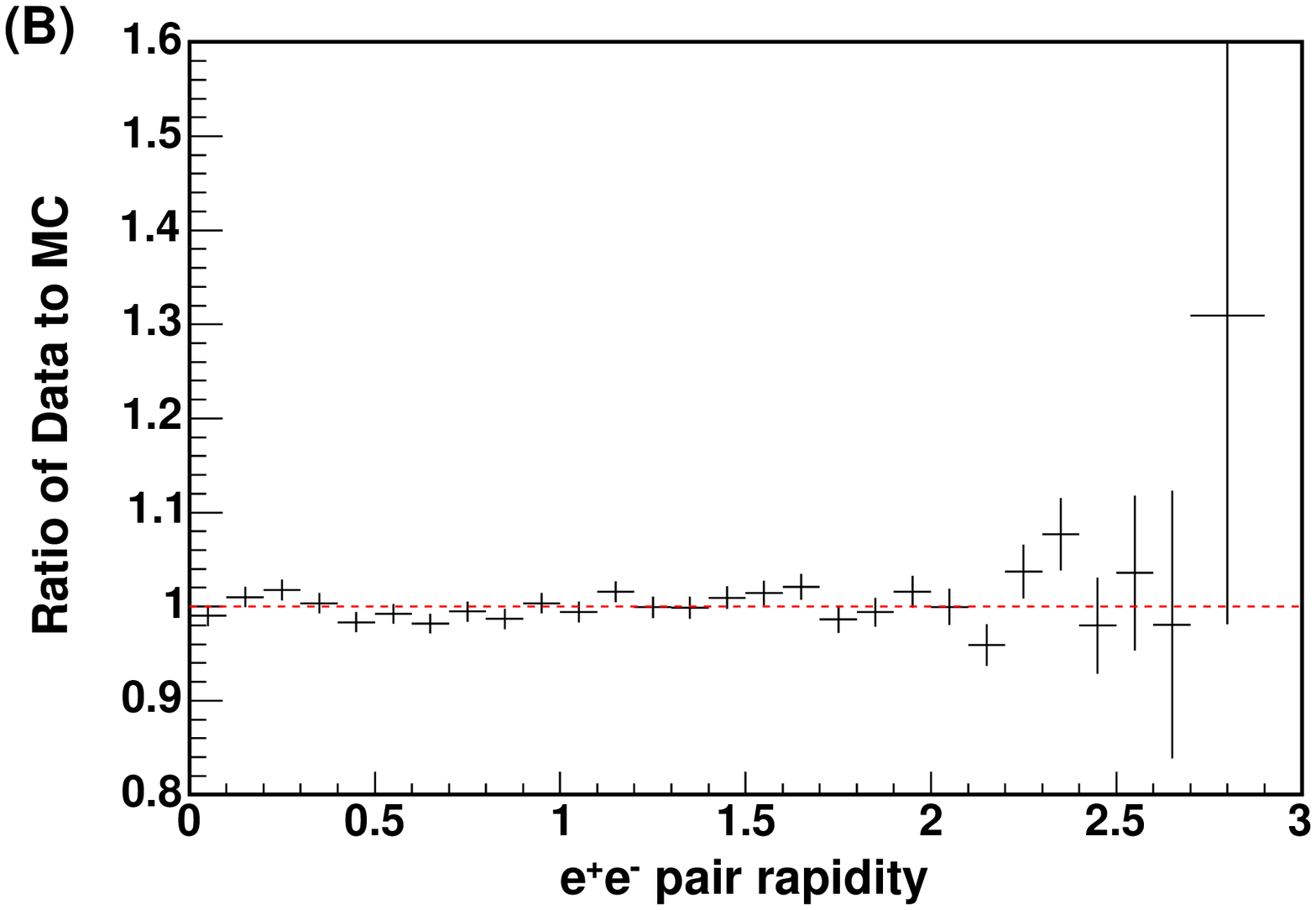}
\end{center}
\caption{The  ratio of the rapidity distributions $dN/dy$  of 
  reconstructed  $e^{+}e^{-}$ pairs in data to
  the results from the simulation using 
  {\sc pythia}, CTEQ5L LO PDFs and CDF W/Z tuning parameters~\cite{pythia}.
(A)   {\sc pythia} before modification of the rapidity distribution.
(B)   {\sc pythia}  after modification of the rapidity distribution .
}
\label{fig:yratio}
\end{figure}

\begin{figure}[ht]
\begin{center}
 \includegraphics[width=3.0in]{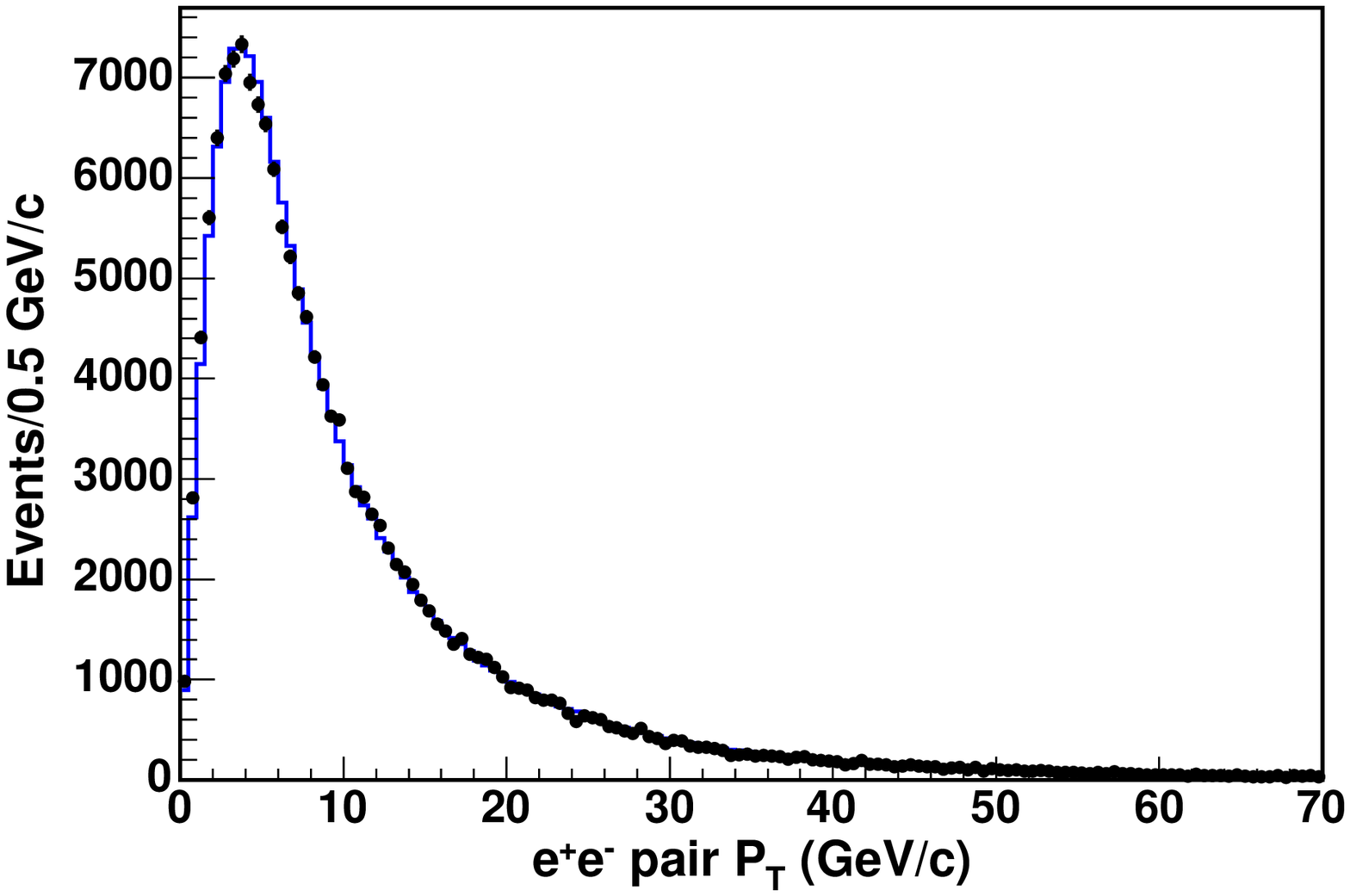}
\end{center}
\caption{The reconstructed  transverse momentum distribution  of $e^{+}e^{-}$ pairs
(after all event selection cuts)  compared to the results from the simulation using 
{\sc pythia}, CTEQ5L LO PDFs and CDF W/Z tuning parameters~\cite{pythia}.
The black points are data and the blue solid line is the {\sc pythia} MC prediction.}
\label{fig:pt}
\end{figure}

\begin{figure}[ht]
\begin{center}
 \includegraphics[width=3.0in]{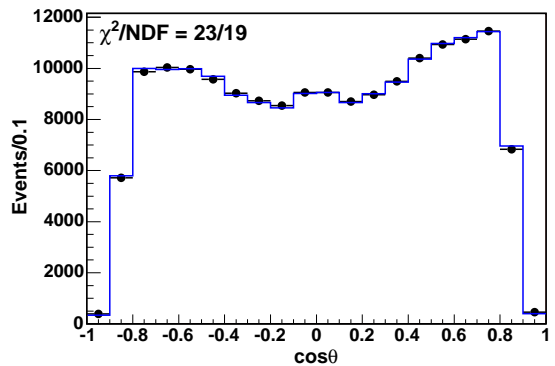}
\end{center}
\caption{The reconstructed polar angle  distribution  of $e^{+}e^{-}$ pairs 
(after all event selection cuts)  compared to the results from the simulation using 
{\sc pythia}, CTEQ5L LO PDFs and CDF W/Z tuning parameters~\cite{pythia}.
The black points are data and the blue solid line is  the {\sc pythia} MC prediction.}
\label{fig:angular}
\end{figure}

 The acceptance ($A$) and efficiencies ($\epsilon$) are determined as a function of boson rapidity.
The contributions of each topology to the product $A\times \epsilon$ are shown in Fig. \ref{fig:ate}.
  
\begin{figure}[ht]
\begin{center}
\includegraphics[width=3.0in]{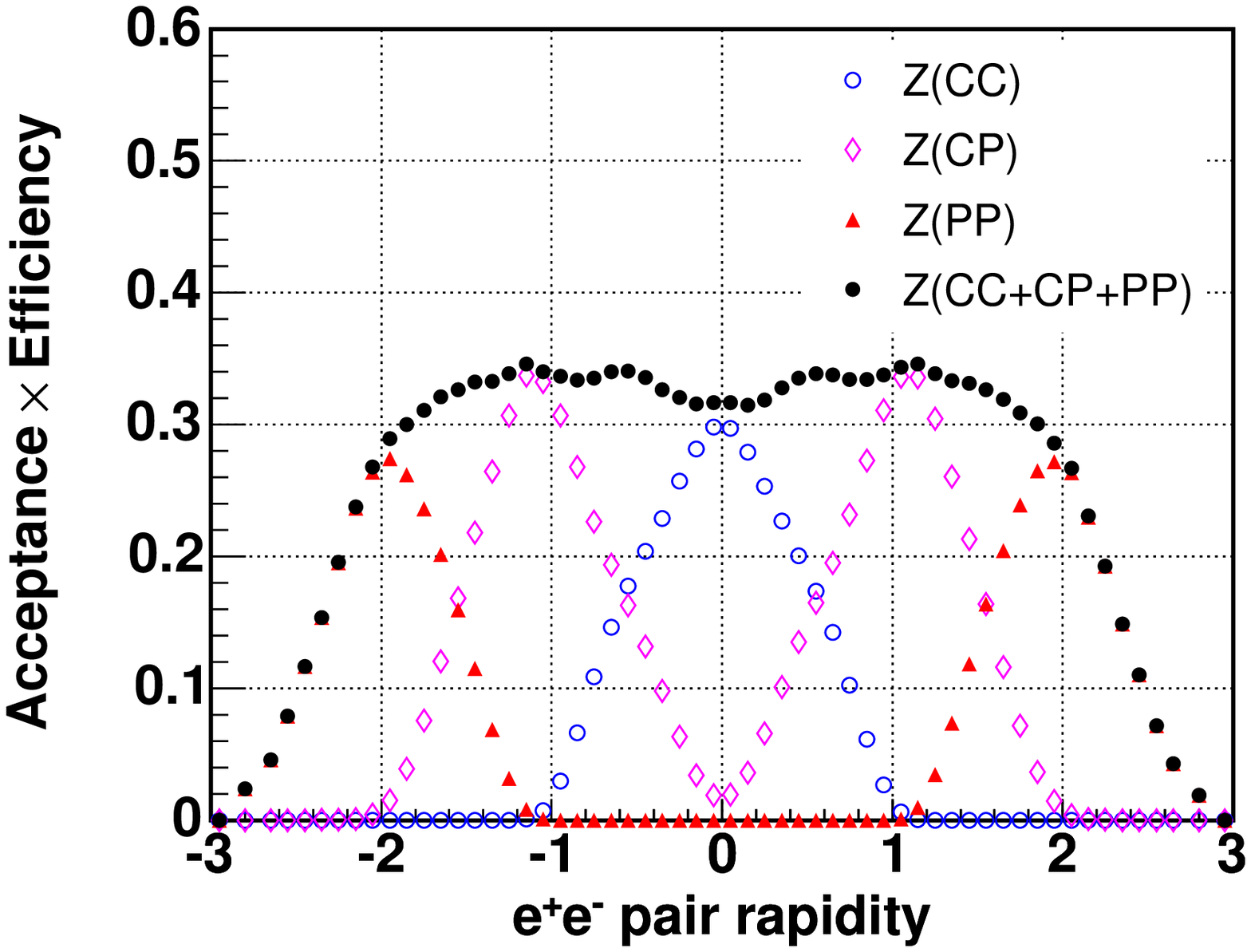}
\end{center}
\caption{
The product of kinematic acceptance and event selection efficiency $vs.$ the rapidity of the $e^{+}e^{-}$ pair.
The black points are the sum of all topologies.}
\label{fig:ate}
\end{figure}

\section{Differential and
Total Cross Sections} \label{sec:measurement}

The differential cross section is given by
   \begin{eqnarray*}
    \small
    \frac{d\sigma(\gamma^*/Z)}{dy}(y) = \frac{N_{sig}(y) - N_{bkg}(y)}{ c(y)\Delta y\epsilon_{zvtx} \Sigma_{i}[
    (A_i\times \epsilon_i (y))\epsilon^{i}_{trig}(y)\mathcal{L}_{i}]}
  \end{eqnarray*} 
  where $N_{sig}(y) - N_{bkg}(y)$ is the number of events after subtracting background,
  $c(y)$ is a correction factor used in order to yield $d\sigma/dy$ at the center of the bin,
 and $\Delta y$ is the $y$ bin size ($\Delta y = 0.1$ up to $y=2.7$ and $\Delta y = 0.2$ for the last bin, $2.7<y\leq 2.9$).
 The sum index $i$ runs over the $e^{+}e^{-}$  topologies (CC, CP, PP), $A_i\times \epsilon_i(y)$ 
 is the combined acceptance and event selection efficiency, $\epsilon^{i}_{trig}(y)$
 is the trigger efficiency,  
 $\mathcal{L}_{i}$ is the total integrated luminosity for each topology, and
 $\epsilon_{zvtx}$ is the acceptance for the $p\bar{p}$ collision vertex to occur
  within $z=\pm60$ cm of the center of the detector.
  The $\epsilon_{zvtx}$ in the data taken before June 2006 is $95.8\pm 0.2\%$ and
 after that is $96.8\pm 0.2\%$.

 Systematic uncertainties in  $d\sigma/dy$ originate from uncertainties
 in the  estimates of the acceptance, backgrounds, electron identification efficiency, 
 SVX tracking efficiency, and modeling of material in the detector. 
 Uncertainties associated with correcting the acceptance for differences
  between kinematic distributions in data and simulation are found to be negligible. 
 The total systematic uncertainty is
  $\sim1.0\%$ of $d\sigma/dy$ for $|y|<2.5$,  increasing
   to $10.0\%$ at $|y|=2.9$. The uncertainty on the integrated luminosity (lum.) is $6\%$. 
 
\begin{figure}[ht]
\begin{center}
\includegraphics[width=3.0in]{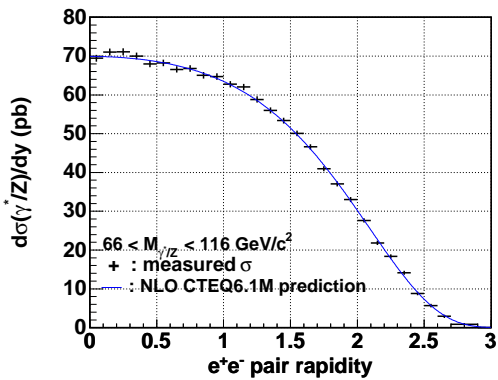}
\end{center}
\caption{
The measured $d\sigma/dy$ for $p\bar{p}\to Z^{0}/\gamma^*\to e^+e^-$ over the entire rapidity range.
The points are the measured cross section versus $|y|$ and the solid line is the theory prediction
(scaled to the measured total cross section)  for CTEQ6.1M(NLO) PDFs.
}
\label{fig:ew}
\end{figure}

\begin{figure}[ht]
\begin{center}
\includegraphics[width=3.5in]{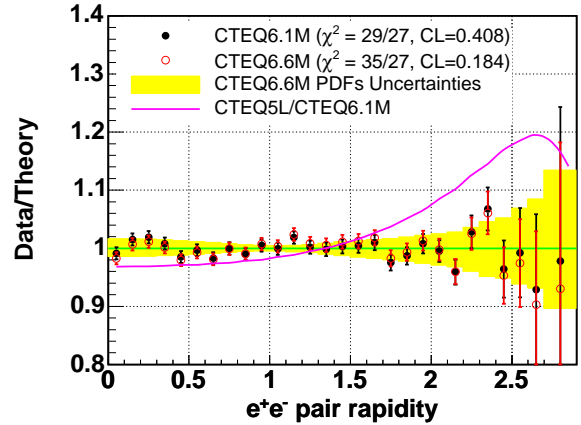}
\end{center}
\caption{The ratio of the experimental distribution of $d\sigma/dy$ (statistical and  systematic uncertainties combined) 
to the theoretical predictions with CTEQ NLO PDFs (CTEQ6.1M and CTEQ6.6M). 
 The yellow bands corresponds to the CTEQ6.6M PDFs  $90\%$ C.L. uncertainties. The $\chi^{2}$ test includes
 the data statistical and systematical uncertainties.}
\label{fig:frac4}
\end{figure}
\begin{figure}[ht]
\begin{center}
\includegraphics[width=3.5in]{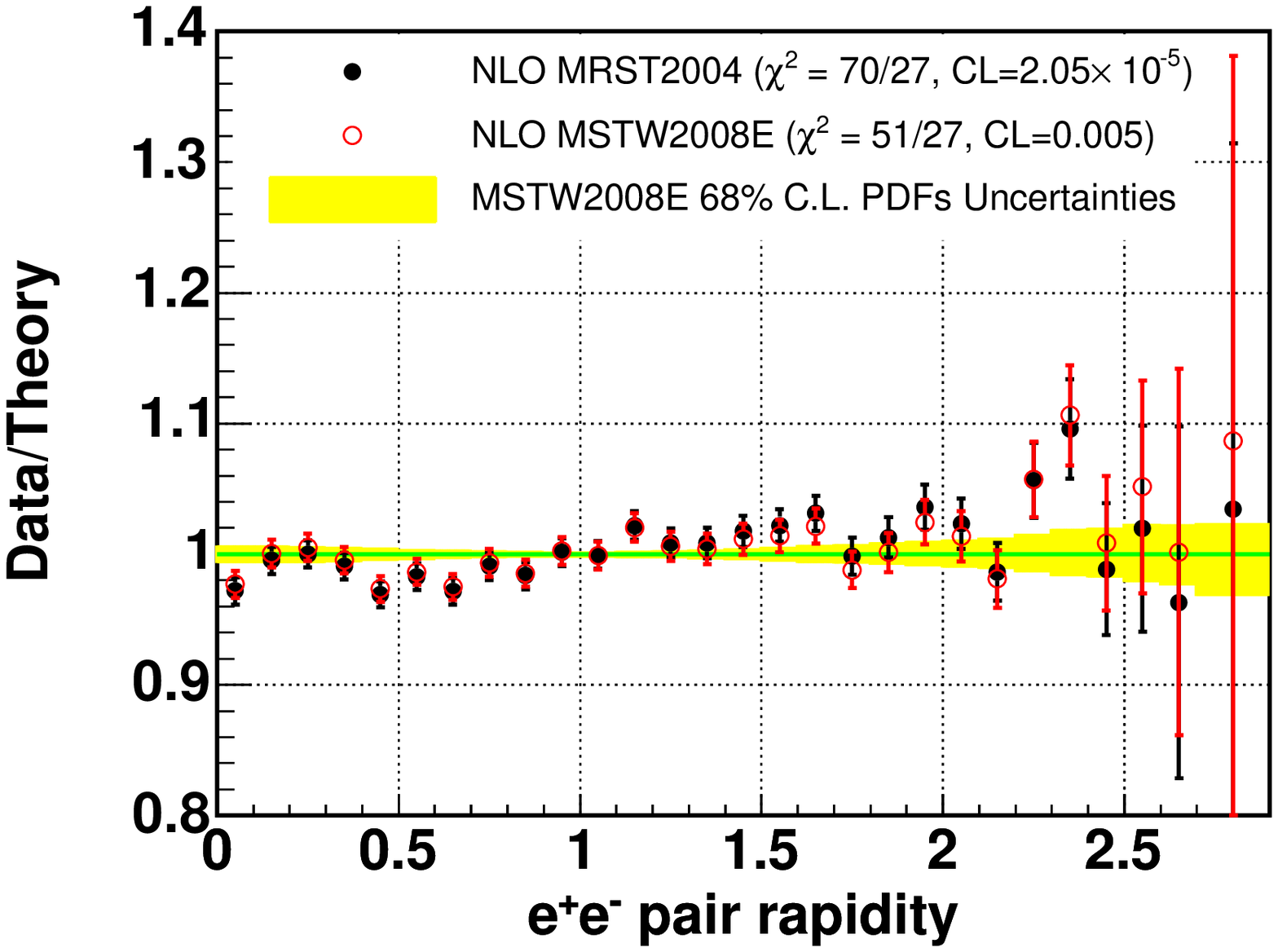}
\end{center}
\caption{The ratio of the experimental distribution of $d\sigma/dy$ (statistical and  systematic uncertainties combined) 
to the theoretical predictions with MRST2004  and MSTW2008E  NLO PDFs.
 The yellow bands corresponds to the MSTW2008E PDFs  $68\%$ C.L. uncertainties.
 The $\chi^{2}$ test includes  the data statistical and systematical uncertainties.}
\label{fig:frac5}
\end{figure}
\begin{figure}[ht]
\begin{center}
\includegraphics[width=3.5in]{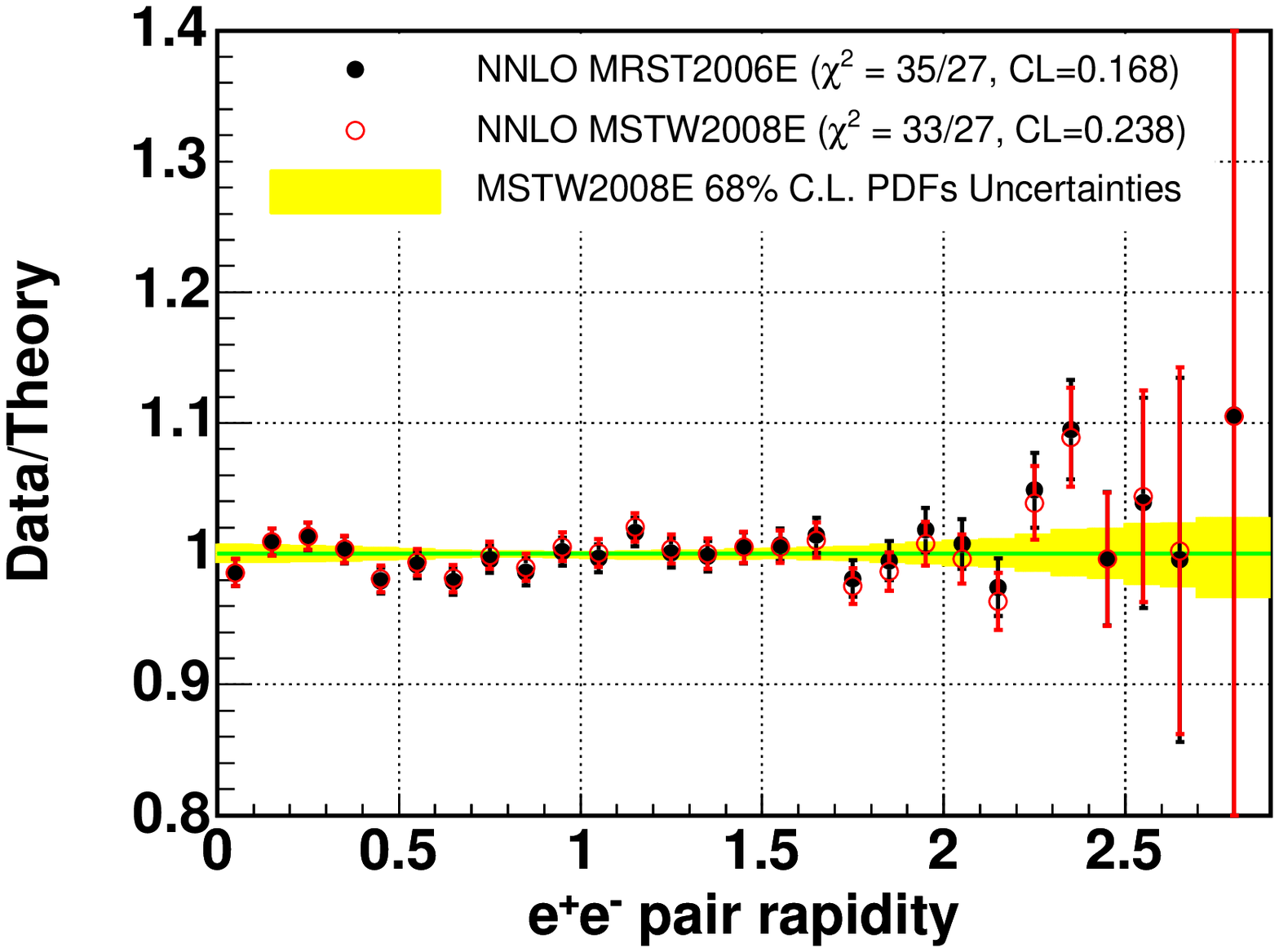}
\end{center}
\caption{The ratio of the experimental distribution of $d\sigma/dy$ (statistical and  systematic uncertainties combined) 
to the theoretical predictions with  MRST2006E and MSTW2008E NNLO PDFs.
 The yellow bands corresponds to the MSTW2008E PDFs $68\%$ C.L. uncertainties.
 The $\chi^{2}$ test includes  the data statistical and systematical uncertainties.}
\label{fig:frac6}
\end{figure}

\begin{table}[ht]
\caption{\label{tab:fonts}
Differential cross sections for production of $e^{+}e^{-}$ pairs in the mass range 
 $66<M_{ee}<116$ GeV/c$^{2}$.
The first and second uncertainties are statistical and systematical, respectively. 
The $6\%$ luminosity uncertainty is not included. The  quoted  $y$  values correspond to the center of the bin. 
The bin size is 0.1 up to $y=2.7$ and 0.2 for the last bin.}
\begin{ruledtabular}
\begin{tabular}{lclc}
$~~y$  &  $d\sigma/dy[pb]~~$  & $~~y$ & $d\sigma/dy[pb]~~$ \\ \hline
0.05  & $69.46 \pm 0.73 \pm 0.49$ & 1.55 & $50.07 \pm 0.62 \pm 0.37$ \\
0.15  & $71.03 \pm 0.74 \pm 0.49$ & 1.65 & $46.59 \pm 0.61 \pm 0.35$ \\
0.25  & $71.10 \pm 0.74 \pm 0.49$ & 1.75 & $40.97 \pm 0.58 \pm 0.34$ \\
0.35  & $70.01 \pm 0.72 \pm 0.48$ & 1.85 & $37.04 \pm 0.56 \pm 0.33$ \\
0.45  & $67.97 \pm 0.70 \pm 0.47$ & 1.95 & $33.02 \pm 0.55 \pm 0.31$ \\
0.55  & $68.22 \pm 0.70 \pm 0.47$ & 2.05 & $27.65 \pm 0.52 \pm 0.25$ \\
0.65  & $66.58 \pm 0.69 \pm 0.47$ & 2.15 & $21.84 \pm 0.49 \pm 0.23$ \\
0.75  & $66.81 \pm 0.70 \pm 0.48$ & 2.25 & $18.35 \pm 0.50 \pm 0.20$ \\
0.85  & $65.05 \pm 0.69 \pm 0.49$ & 2.35 & $14.13 \pm 0.49 \pm 0.17$ \\
0.95  & $64.70 \pm 0.69 \pm 0.50$ & 2.45 & $8.80 \pm 0.45 \pm 0.10$ \\
1.05  & $62.74 \pm 0.67 \pm 0.50$ & 2.55 & $5.68 \pm 0.44 \pm 0.09$ \\
1.15  & $62.02 \pm 0.66 \pm 0.49$ & 2.65 & $2.93 \pm 0.41 \pm 0.15$ \\
1.25  & $58.80 \pm 0.65 \pm 0.48$ & 2.80 & $0.87 \pm 0.22 \pm 0.11$ \\
1.35  & $56.02 \pm 0.65 \pm 0.43$ & 2.95 & $-$ \\
1.45  & $53.37 \pm 0.63 \pm 0.40$ &            \\ 
\end{tabular}
\end{ruledtabular}
\end{table}
\begin{table}[ht]
\caption{\label{tab:cross} A comparison of the measured  total cross section
for the production of 
 $e^{+}e^{-}$ pairs in the mass range $66<M_{ee}<116$ GeV/c$^{2}$  to theory calculations. 
}
\begin{ruledtabular}
\begin{tabular}{lc}
$Model$  &  $Total~cross~section $   \\ \hline
CTEQ5L(LO)   &  $183.3$  \\
MRST2001E(NLO)   & $241.0^{+2.8}_{-3.4}$ \\
MRST2004(NLO)   & $241.2$ \\
MSTW2008E(NLO)   & $242.6^{+4.6}_{-5.5}$ \\
CTEQ6.1M(NLO)   & $236.1^{+9.3}_{-9.2}$ \\
CTEQ6.6M(NLO)   & $238.7^{+7.1}_{-7.0}$ \\
MRST2006E(NNLO)   & $251.6^{+2.8}_{-3.1}$ \\
MSTW2008E(NNLO)   & $248.7^{+5.1}_{-4.0}$ \\
Data   & $256.6   \pm 0.7 \pm 2.0  \pm 15.4$ \\
\end{tabular}
\end{ruledtabular}
\end{table}
\begin{table}[ht]
\caption{\label{tab:diff} 
A comparison of the shape of the
measured $d\sigma/dy$ distribution  to theoretical predictions with several choices of PDFs.
The theoretical  distributions
are normalized to the measured total cross section of $256.6$ pb. The   $\chi^{2}$  for
27 degrees of freedom includes  statistical and systematic uncertainties.}
\begin{ruledtabular}
\begin{tabular}{lcc}
$Model$  &  $\chi^{2}/DOF$ &$CL$  \\ \hline
CTEQ5L(LO)   &  $ 242/27 $ & $-$   \\
MRST2001E(NLO)   & $76/27 $ &  $ 2.908\times 10^{-6} $ \\
MRST2004(NLO)   & $70/27 $ &  $ 2.049\times 10^{-5} $ \\
MSTW2008E(NLO)   & $51/27$ & $ 0.005 $ \\
CTEQ6.1M(NLO)   & $29/27 $ &  $ 0.408 $ \\
CTEQ6.6M(NLO)   & $35/27 $ &  $ 0.184 $ \\
MRST2006E(NNLO)    & $35/27$ & $ 0.168 $ \\
MSTW2008E(NNLO)   &  $33/27 $ &  $ 0.238 $  \\
\end{tabular}
\end{ruledtabular}
\end{table}

\section{Results} \label{sec:result}

 The measured $d\sigma/dy$ values, which are symmetric about $y=0$, 
are shown versus  $|y|$, with  statistical and  systematic uncertainties,  in  Fig. \ref{fig:ew} and  Table \ref{tab:fonts}. 
 The total cross section,  derived from integrating $d\sigma/dy$ 
up to $|y|=2.9$,  is $\sigma = 256.6\pm0.7({\rm stat.})\pm2.0({\rm syst.})\pm15.4({\rm lum.})$ pb.
These results are compared to QCD predictions at LO with CTEQ5L~\cite {cteq5L}, at  NLO \cite {nlo_ms} with MRST2001 (NLO) \cite{mrst_nlo_theory},   MRST2004 (NLO) \cite{mrst2004_nlo_theory}, \textsc{CTEQ6.1M} (NLO) \cite{cteq_theory},
 \textsc{CTEQ6.6M} (NLO) \cite{cteq66_theory}, and MSTW2008 (NLO) \cite{mrst2008pdfs}
 PDFs, and  at NNLO \cite {nnlo_ms}  with  
 \textsc{MRST2006E} (NNLO) \cite{mrst_nnlo_theory} and MSTW2008 (NNLO) \cite{mrst2008pdfs} PDFs.
The measured total cross section 
is consistent with both  NLO and NNLO calculations as shown in Table \ref{tab:cross}. 

In comparing the shape of the measured $d\sigma/dy$ to theory, the latter distributions
are normalized to the measured total cross section of $256.6$ pb.
The ratios of the measured $d\sigma/dy$  to the QCD  calculations  at  LO, NLO and NNLO
with the above mentioned PDFs are  shown in Figures \ref{fig:frac4}, \ref{fig:frac5}, and \ref{fig:frac6}.
The yellow bands in the figures correspond to the uncertainties associated to the MSTW2008E (NLO and NNLO) PDFs, 
which are given with $68\%$ C.L. errors and to the those associated to the other sets of PDFs, given with $90\%$ C.L. errors.
A  $\chi^{2}$ comparison (including  statistical and systematic uncertainties)  is shown in Table \ref{tab:diff}.
Better agreement is obtained for the MSTW (2008) PDF at NNLO compared to NLO. The NLO CTEQ6.1M, CTEQ6.6M, the NNLO MRST2006 and MSTW2008 PDFs all describe the data well. The older NLO MRST (2004) set provides a poorer description of the data and, as expected, so does the LO PDF, CTEQ5L. The MSTW(2008) PDF used a preliminary version of the data presented in
this paper in their fit.
The correlations \cite{jyhan_thesis,dsigmady_sys} between the uncertainties in different $y$ bins are included in the $\chi^{2}$ comparison.

  In summary, high-statistics measurement of $\gamma^{*}/Z$  production in the $Z$
   mass region and of its rapidity distribution are found to agree with theoretical calculations that use recent NLO and NNLO PDFs. (A preliminary version of these results has been used in the determination of the most recent PDFs). 
  The precise measurement of  the total production cross section,  $\sigma = 256.6\pm0.7({\rm stat.})\pm2.0({\rm syst.})\pm15.4({\rm lum.})$ pb  can be used to set the normalization of other processes at the Tevatron and the LHC.
  
\begin{acknowledgments}
We thank the Fermilab staff and the technical staffs of the participating institutions for their vital contributions. This work was supported by the U.S. Department of Energy and National Science Foundation; the Italian Istituto Nazionale di Fisica Nucleare; the Ministry of Education, Culture, Sports, Science and Technology of Japan; the Natural Sciences and Engineering Research Council of Canada; the National Science Council of the Republic of China; the Swiss National Science Foundation; the A.P. Sloan Foundation; the Bundesministerium f\"ur Bildung und Forschung, Germany; the World Class University Program, the National Research Foundation of Korea; the Science and Technology Facilities Council and the Royal Society, UK; the Institut National de Physique Nucleaire et Physique des Particules/CNRS; the Russian Foundation for Basic Research; the Ministerio de Ciencia e Innovaci\'{o}n, and Programa Consolider-Ingenio 2010, Spain; the Slovak R\&D Agency; and the Academy of Finland. We also thank Robert Thorne for valuable discussions.
\end{acknowledgments}


\begin{thebibliography}{99}
\bibitem{cdf_coord}
CDF coordinates are ($\theta$,$\phi$,$z$), where $\theta$ is the polar angle relative to
the proton beam (the $+z$ axis), and $\phi$ the azimuth.
The pseudorapidity is $\eta=-ln~\tan(\theta/2)$. For an $e^{+}+e^{-}$ pair
 $P_{T}=P\sin\theta$,  $E_{T}=E\sin\theta$, 
 and $y=\frac{1}{2}ln\frac{E+P_{z}}{E-P_{z}}$, where
$P$ and $P_{z}$ are the magnitude and z component of the momentum,
and E is the energy of the $e^{+}+e^{-}$ pair.
\end{thebibliography}
\end{document}